# K-Reach: Who is in Your Small World

James Cheng
Nanyang Technological
University, Singapore
j.cheng@acm.org

Zechao Shang
The Chinese University of
Hong Kong
zcshang@se.cuhk.edu.hk

Hong Cheng
The Chinese University of
Hong Kong
hcheng@se.cuhk.edu.hk

Haixun Wang
Microsoft Research Asia
Beijing, China
haixunw@microsoft.com

Jeffrey Xu Yu
The Chinese University of
Hong Kong
yu@se.cuhk.edu.hk

## ABSTRACT

We study the problem of answering **$k$-hop reachability queries** in a directed graph, i.e., whether there exists a directed path of length $k$, from a source query vertex to a target query vertex in the input graph. The problem of $k$-hop reachability is a general problem of the classic reachability (where $k = \infty$). Existing indexes for processing classic reachability queries, as well as for processing shortest path queries, are not applicable or not efficient for processing $k$-hop reachability queries. We propose an index for processing $k$-hop reachability queries, which is simple in design and efficient to construct. Our experimental results on a wide range of real datasets show that our index is more efficient than the state-of-the-art indexes even for processing classic reachability queries, for which these indexes are primarily designed. We also show that our index is efficient in answering $k$-hop reachability queries.

## 1. INTRODUCTION

The **reachability query**, which asks whether one vertex can reach another vertex in a directed and unweighted graph, is a basic operator for a variety of databases (e.g., XML, RDF) and network applications (e.g., social and biological networks). The problem of how to efficiently answer reachability queries has attracted a lot of interest lately [2, 3, 5, 6, 7, 14, 15, 16, 17, 19, 23, 24, 26, 27, 28, 29, 32, 34]. There are also several extensions to the classic reachability problem; for example, reachability in uncertain graphs where the existence of an edge is given by a probability [21], and reachability with constraints such as edges on the path must have certain labels [20], etc.

In this work, we study a new type of reachability queries. Instead of asking whether a vertex $t$ is reachable from a vertex $s$, we ask whether $t$ is reachable within $k$ hops from $s$. In other words, *the query asks whether there exists a path from $s$ to $t$ such that the length of the path is no more than $k$*. We call this problem the **$k$-hop reachability** problem.

The primary motivation for the $k$-hop reachability problem is that *in many real life networks (e.g., wireless or sensor networks, the Web and Internet, telecommunication networks, social networks, etc.), the number of hops within which $s$ can reach $t$ indicates the level of influence $s$ has over $t$*. Applications on such networks can benefit more from $k$-hop reachability than classic reachability (i.e., $k = \infty$). We give some examples as follows.

In a wireless or sensor network, where a broadcasted message may get lost during any hop, the probability of reception degrades exponentially over multiple hops. In these applications, reachability may not be meaningful or have much practical use, while $k$-hop reachability, since it can model the level and sphere of the influence, is helpful in many analytical tasks.

In many real life networks, the $k$-hop reachability between two vertices is of more interest when the value of $k$ is small. In a sensor network, the probability of reception degrades exponentially. Thus, after a few hops, the probability will be well below the threshold of interest. In social networks, although there is a well-known six-degrees-of-separation theory (i.e., any two persons are only 6 or fewer hops away from each other), the degree of acquaintance may even decrease super-exponentially (i.e., two persons may hardly know each other if they are just 3 hops apart).

Clearly, we are more interested to know if two persons are connected within only a few hops, instead of beyond 6 hops for which case one can almost know for sure they are connected. On the other hand, a small $k$ does not necessarily make the problem easier. Consider an application that, given two persons, asks whether one is reachable from the other within 6 hops. A naive implementation is to invoke a breadth-first search (BFS). However, among the search quests in reality, a majority of them have at least one of the persons as a celebrity at some level of the BFS. A BFS from a celebrity (e.g., Lady Gaga, who has 40,000,000 fans on Facebook) can quickly cover a considerable proportion of the entire social network within as small as 3 hops, and is clearly out of the question for online query processing.

The problem of $k$-hop reachability cannot be derived from classic reachability, which is actually a special case of $k$-hop reachability, i.e., when $k = \infty$. Indeed, the $k$-hop problem is more challenging since more information is required to answer the $k$-hop reachability query. To see this, consider the *transitive closure* [26] of the adjacency matrix of a directed graph. If we are given this transitive matrix, we can find out if $s$ can reach $t$ instantly by checking whether their corresponding entry in the transitive matrix is 1 or 0. However, for large graphs, it is infeasible to pre-compute and store the transitive closure, as it takes $O(n^2)$ space, where $n$ is the number of vertices in the graph. Thus, the reachability problem is essentially the problem of how to effectively "encode" the 0-1 transitive matrix into a small index structure which still provides efficient lookup of reachability between any two vertices. For $k$-hop reachability, the matrix we need to "encode" is no longer a 0-1 matrix. Instead, each entry $c_{st}$ in the matrix contains the length of the shortest path from $s$ to $t$. Clearly, this matrix contains much more information than the 0-1 transitive matrix. In Section 3, we further analyze in details why all the existing indexes are not suitable for processing $k$-hop reachability queries.

We propose an efficient index, called **$k$-reach**, to process $k$-hop reachability queries. The $k$-reach index is constructed based on the concept of *vertex cover* [18]. *The main idea of the index design is based on the fact that all vertices in a graph are reachable within 1 hop of some vertex in the vertex cover of the graph.* The vertex cover is small for a wide range of real world graphs. Thus, we only need to pre-compute $k$-hop reachability information among a small portion of the vertices, while we also show that it only requires at most 2 bits for encoding each $k$-hop reachability information.

Another advantage of $k$-reach is that it allows the inclusion of all high-degree vertices in the vertex cover. This not only gives the same coverage with a smaller vertex cover (and hence reduces the index size), but also allows queries that involve high degree vertices to be answered more efficiently (e.g., the "Lady Gaga" example given earlier). In addition, we also propose a method that further reduces the size of the index by extending the coverage of the vertex cover to $h$-hop neighbors.

The $k$-reach index is simple in design and easy to implement. The index can handle both classic reachability queries and $k$-hop reachability queries. We conducted experiments on a wide range of 15 real datasets. Our results show that even for processing classic reachability queries, $k$-reach is significantly more efficient than the state-of-the-art indexes [23, 24, 28, 32] that are primarily designed for classic reachability. We also show that $k$-reach is orders of magnitude faster than $k$-hop BFS and the shortest-path distance index [13], which can be used to answer $k$-hop reachability queries, thus demonstrating the need for a $k$-hop reachability index. In addition, the performance of $k$-reach is stable for both small and large values of $k$. The results also show that $k$-reach is efficient to construct and has a small storage size.

**Organization.** Section 2 formally defines the notations and the problem. Section 3 analyzes the difficulties of applying the existing works for handling $k$-hop reachability and highlight the challenges. Section 4 presents the details of the $k$-reach index. Section 5 describes a method to further reduce the index size. Section 6 reports the experimental results. Section 7 discusses other related works, followed by the conclusions in Section 8.

## 2. PROBLEM DEFINITION

Table 1 lists the notations that are frequently used in this paper.

Let $G = (V, E)$ be an unweighted, directed graph, where $V$ is the set of vertices and $E$ is the set of edges in $G$, respectively. An edge $(u, v) \in E$ is a directed edge from $u$ to $v$, while $(v, u) \in E$ means that the edge is directed from $v$ to $u$.

Given a pair of vertices $s$ and $t$ in $G$, we say that $t$ is **reachable** from $s$, denoted by $s \to t$, if there exists a simple directed path $P = \langle s, \cdots, t \rangle$ in $G$. If $|P| \le k$, where $|P|$ is the path length (i.e., the number of edges on $P$), then $t$ is also **$k$-hop reachable** from $s$, denoted by $s \to_k t$. A **reachability query** is to determine whether $s \to t$, while a **$k$-hop reachability query** is to determine whether $s \to_k t$. Note that a reachability query is in fact an $n$-hop or $\infty$-hop reachability query.

**The $k$-hop reachability indexing problem.** *Given an unweighted,*

**Table 1: Frequently used notations**

| Notation | Description |
|---|---|
| $G = (V, E)$ | A directed, unweighted graph |
| $n$ or $m$ | The number of vertices or edges in $G$ |
| $s \to t$ | $t$ is reachable from $s$ |
| $s \to_k t$ | $t$ is reachable from $s$ within $k$ hops |
| $inNei(v, G)$ | the set of in-neighbors of $v$ in $G$ |
| $inDeg(v, G)$ | the in-degree of $v$ in $G$, i.e., $|inNei(v, G)|$ |
| $outNei(v, G)$ | the set of out-neighbors of $v$ in $G$ |
| $outDeg(v, G)$ | the out-degree of $v$ in $G$, i.e., $|outNei(v, G)|$ |
| $Nei(v, G)$ | the set of neighbors of $v$ in $G$ |
| $Deg(v, G)$ | the out-degree of $v$ in $G$, i.e., $|Nei(v, G)|$ |

*directed graph $G$, the paper proposes an index structure for $G$ to answer $k$-hop reachability queries.*

The following notations will also be used throughout the paper in the discussion of our indexing and query processing algorithms.

Given $G = (V, E)$, define $n = |V|$ and $m = |E|$. We denote the set of **in-neighbors** of a vertex $v$ in $G$ by $inNei(v, G) = \{u : (u, v) \in E\}$, and the **in-degree** of $v$ in $G$ as $inDeg(v, G) = |inNei(v, G)|$. Similarly, we denote the set of **out-neighbors** of $v$ in $G$ by $outNei(v, G) = \{u : (v, u) \in E\}$, and the **out-degree** of $v$ in $G$ as $outDeg(v, G) = |outNei(v, G)|$. We also denote the set of **neighbors** of $v$ in $G$ by $Nei(v, G) = (inNei(v, G) \cup ourNei(v, G))$, and the **degree** of $v$ in $G$ as $Deg(v, G) = |Nei(v, G)|$.

## 3. REACHABILITY VS. K-HOP REACHABILITY

In this section, we analyze the suitability of the existing graph reachability indexes [2, 3, 5, 6, 7, 14, 15, 16, 17, 19, 23, 24, 26, 27, 28, 29, 32, 34] for processing $k$-hop reachability queries. An understanding of these existing works with their relation to the problem of $k$-hop reachability query processing helps not only the work in this paper but also potential future work along this direction.

We categorize the existing works of reachability indexing into six approaches and then show that they cannot be applied or are inefficient for processing $k$-hop reachability queries. Note that some existing works may fall into more than one category since they may combine different approaches to solve the problem.

### 3.1 Directed Acyclic Graph based Approach

The first category of indexes is related to *directed acyclic graph* (**DAG**). Many existing indexes for processing reachability queries assume that the input graph is a DAG [5, 6, 7, 19, 23, 24, 26, 28, 34], because if the input graph is not a DAG, one can pre-process it and turn it into a DAG as follows. First, compute all the *strongly connected components* (**SCCs**) in the input graph. Then, condense each SCC into a single super-vertex, where each super-vertex is a vertex in the DAG. Finally, a directed edge $(c_1, c_2)$ is added from a vertex $c_1$ to another vertex $c_2$ in the DAG iff there exists a directed edge $(u, v)$ in the original graph such that $u$ is in $c_1$ and $v$ is in $c_2$ (note that $c_1$ and $c_2$ are two SCCs in the original graph).

Condensing a general graph into a DAG can save space and it works for processing reachability queries since all vertices within an SCC are pairwise reachable from each other. However, for processing $k$-hop reachability queries, the DAG-based approach fails because two reachable vertices in the DAG may not be $k$-hop reachable in the original graph, since the shortest path connecting them may have been condensed into a much shorter path (of length $\le k$) in the DAG. To answer a $k$-hop reachability query, one has to expand the vertices involved in the DAG to their corresponding SCCs in the original graph in order to examine the $k$-hop information, which is no cheaper than directly checking $k$-hop in the original graph.

## 3.2 Traversal-based Vertex Coding Approach

The second category of graph reachability indexes focuses on designing some *vertex coding scheme based on graph traversal* [2, 5, 7, 27, 29, 32, 34]. A traversal (e.g., DFS) of a graph assigns each vertex a pair of codes according to the traversal order (e.g., the discovery time and finish time of a vertex in a DFS). The pair of codes obtained from a traversal forms an interval, which can be further modified to capture more information of descendants or of other relevant links. Then, reachability queries can be answered based on the containment relationship of the intervals. Different graph traversal methods may be applied and there can be multiple traversals depending on the design of the index.

For processing $k$-hop reachability queries, however, the interval containment test of a traversal-based approach fails to capture the $k$-hop requirement. To examine the number of hops from the source vertex to the target vertex, one needs to explore the input graph. Although the vertex coding may help guide the exploration, the process can be as expensive as a trivial BFS to process the $k$-hop reachability query starting from the source vertex.

## 3.3 Chain Cover based Approach

The third category of graph reachability indexes are constructed based on a *chain cover* of the input graph or partially relied on some chain cover [6, 7, 19, 23, 24]. A chain cover of a graph $G = (V, E)$ consists a set of chains, $\{C_1, \cdots, C_t\}$, where $C_i \subseteq V$, $\bigcup_{1 \leq i \leq t} C_i = V$ and $(C_i \cap C_j) = \emptyset$, for $1 \leq i, j \leq t$ and $i \neq j$. For each chain $C_i = \{v_1, \cdots, v_{c_i}\}$, we have $v_x \to v_y$ for $1 \leq x < y \leq c_i$. After computing a chain cover of $G$, each vertex $v \in V$ is assigned a list of chain codes $\{\sigma_1, \cdots, \sigma_t\}$, where $\sigma_i$ indicates that $v$ can reach the vertex at the $\sigma_i$-th position in the chain $C_i$. Thus, a reachability query can be answered by examining the lists of chain codes of the vertices involved.

Since a chain or the list of chain codes of a vertex retain only the reachability information between the vertices, the chain cover based indexes cannot process a $k$-hop reachability query. It is not clear how we may extend the chain cover to contain the information of $k$-hop reachability, since the connections among both the chains and vertices in a chain are all involved. Even though the information of $k$-hop reachability can be indexed in the chain cover, resolving the inter-connection between chains and intra-connection within a chain to process $k$-hop reachability can be complicated and expensive.

## 3.4 2-Hop Cover based Approach

The fourth category of works construct reachability indexes based on the concept of *2-hop cover* [3, 14, 15, 16, 17, 23]. The 2-hop cover approach computes for each vertex $v$ in an input graph $G = (V, E)$ two vertex subsets, $L_{in}(v)$ and $L_{out}(v)$, where $L_{in}(v)$ consists of a set of vertices in $G$ that can reach $v$ and $L_{out}(v)$ consists of a set of vertices in $G$ that can be reached from $v$. Then, a reachability query is answered as follows: a source vertex $s$ can reach a target vertex $t$ if and only if $(L_{out}(s) \cap L_{in}(t)) \neq \emptyset$.

The 2-hop cover clearly also cannot be used to process $k$-hop reachability queries because all distance information between the vertices is lost. The 2-hop cover can be extended to encode the distance information of each reachable vertex in $L_{in}(v)$ or $L_{out}(v)$ related to $v$. However, as shown in many existing works of graph reachability, the 2-hop cover has not only a higher complexity but is also significantly less efficient than the recent indexes in real performance for processing reachability queries, not to mention for processing $k$-hop reachability queries. On the contrary, we show that our approach is efficient for processing both reachability queries and $k$-hop reachability queries.

## 3.5 Shortest-Path Approaches

Indexes for processing shortest-path or distance queries can be trivially used to process $k$-hop reachability queries. Shortest-path or distance query processing, however, has a significantly higher cost than $k$-hop reachability query processing. In particular, the 2-hop cover based indexes [13, 17] are not efficient enough for processing $k$-hop reachability queries, as explained in Section 3.4 and also to be shown in Section 6.3.1. Moreover, the works by Xiao et al. [31] and Wei [30] are designed for undirected graphs, while graph reachability mostly considers directed graphs.

Apart from the above-mentioned indexes, there are also many indexes developed for processing shortest-path and distance queries in planar graphs or road networks (see [1] and the references therein). However, these indexes are specifically optimized for road networks and cannot be applied to directed general graphs.

## 3.6 Other Approaches

Other approaches such as transitive closure [26] can also be extended to encode the $k$-hop reachability information. However, the transitive closure is in general too large to be practical. A recent work has been proposed to compress the transitive closure using bit vector compression techniques [28], which has shown to be effective for processing reachability queries. However, unlike encoding for graph reachability which requires only boolean indicators, encoding the $k$-hop reachability information not only requires more bits for each entry, but also breaks the continuity of long sequences of '0's and '1's which is crucial for the effectiveness of the bit vector compression techniques [28]. More critically, both transitive closure [26] and compression on transitive closure [28] work only on the much smaller DAG of the input graph, while the DAG-based approach is not applicable for processing $k$-hop reachability queries as discussed in Section 3.1.

## 4. A VERTEX-COVER-BASED INDEX

Having discussed the limitations of the existing indexes for processing $k$-hop reachability queries, in this section we propose an efficient index, called **k-reach**, as a solution.

## 4.1 K-Reach: Index Construction

The $k$-reach index is constructed based on the concept of *vertex cover* [18]. We first discuss how to compute a small vertex cover of a graph $G$. Then, we define the index structure and describe the algorithm that constructs the index.

### 4.1.1 Minimum Vertex Cover Approximation

A set of vertices, $S$, is a vertex cover of a graph $G = (V, E)$ if for every edge $(u, v) \in E$, we have $(\{u, v\} \cap S) \neq \emptyset$. Obviously, $V$ itself is a vertex cover of $G$ but is too large to be used to construct an index. Thus, we want to minimize the size of the vertex cover.

A vertex cover $S$ is called a *minimum vertex cover* of $G$ if $S$ has the smallest size among all vertex covers of $G$. The problem of computing the minimum vertex cover is well-known to be NP-hard [18]. However, there is a polynomial time algorithm for computing a *2-approximate minimum vertex cover*, which is given as follows.

We randomly select an edge $(u, v)$ from $E$, add both $u$ and $v$ to $S$, and then remove $u$ and $v$ from $G$, together with all edges

**Algorithm 1** *Construction of k-reach*

**Input**: a directed graph $G = (V, E)$ and an integer $k$
**Output**: a $k$-reach index of $G$

1. Compute a 2-approximate minimum vertex
    cover, $S$, of $G$;
2. Initialize a weighted, directed graph $I = (V_I, E_I, \omega_I)$;
3. $V_I \leftarrow S$;
4. **for each** $u \in S$ **do**
5.     Compute $S_k(u) = \{v : v \in S, u \rightarrow_k v\}$
        by a $k$-hop BFS from $u$;
6.     **for each** $v \in S_k(u)$ **do**
7.         $E_I \leftarrow (E_I \cup \{(u, v)\})$;
8.         **if**($u \rightarrow_{k-2} v$)
9.             $\omega_I((u, v)) \leftarrow (k-2)$;
10.        **else if**($u \rightarrow_{k-1} v$)
11.            $\omega_I((u, v)) \leftarrow (k-1)$;
12.        **else** /* $(u \rightarrow_k v)$ */
13.            $\omega_I((u, v)) \leftarrow k$;
14. **return** $I = (V_I, E_I, \omega_I)$;

incident on the two vertices. Note that all edges incident on $u$ or $v$, whether in-edges or out-edges, can be removed from $G$ because all these edges are covered by either $u$ or $v$ in $S$. This process is repeated until all edges are removed from $G$.

The above algorithm takes $O(m + n)$ time since every edge is only touched once. Let $C$ be a minimum vertex cover of $G$. Then, for every pair of vertices, $u$ and $v$, selected to be included in $S$ in the above process, either $u$ or $v$ must be in $C$, because otherwise the edge $(u, v)$ is not covered by any vertex in $C$. Thus, we have $|S| \leq 2|C|$. From the analysis, we also see that we may simply ignore the direction of the edges in computing a 2-approximate minimum vertex cover of $G$.

### 4.1.2 Definition of k-Reach and Its Construction

We now define the structure of the $k$-reach index as follows.

*Definition* 1 (K-REACH). *Given a directed graph $G = (V, E)$, a vertex cover $S$ of $G$, and an integer $k$, the **k-reach** index of $G$ is a weighted, directed graph $I = (V_I, E_I, \omega_I)$ defined as follows.*

- $V_I = S$.

- $E_I = \{(u, v) : u, v \in S, u \rightarrow_k v\}$.

- $\omega_I$ *is a weight function that assigns a weight to each edge $e = (u, v) \in E_I$ as follows:*

    - *if $u \rightarrow_{k-2} v$, then $\omega_I(e) = (k-2)$;*
    - *else if $u \rightarrow_{k-1} v$, then $\omega_I(e) = (k-1)$;*
    - *else if $u \rightarrow_k v$, then $\omega_I(e) = k$.*

Note that in Definition 1, "$u \rightarrow_{k-2} v$" implies "$u \rightarrow_{k-1} v$", both of which also imply "$u \rightarrow_k v$".

Next, we describe the index construction process, as shown in Algorithm 1.

Algorithm 1 first computes a 2-approximate minimum vertex cover, $S$, of the input graph $G$ by the algorithm given in Section 4.1.1. Then, it constructs the graph $I = (V_I, E_I, \omega_I)$ by performing a breath-first search (BFS) of $G$ within $k$ hops from each starting vertex $u \in S$. This process computes the set of all vertices in $S$ that can be reached from $u$ in $k$ hops in $G$, i.e., the set $S_k(u)$ in Line

5. The rest of the algorithm is simply including each edge $(u, v)$ in $E_I$, for each $v \in S_k(u)$, and assigning the weight to $(u, v)$.

We give an example of the $k$-reach index constructed by Algorithm 1 as follows.

*Example 1.* Given the graph $G$ in Figure 1. Assume that the 2-approximate algorithm randomly picks the edges, $(b, d)$ and $(g, i)$, in $G$. Then, $\{b, d, g, i\}$ forms the set of 2-approximate minimum vertex cover of $G$. We can verify that $\{b, d, g, i\}$ is indeed a vertex cover of $G$, since every edge in $G$ is incident on at least one of the vertices in $\{b, d, g, i\}$.

Let $k = 3$. The $k$-reach graph, $I = (V_I, E_I, \omega_I)$, of $G$ is shown in Figure 2. Since $k = 3$, the possible edge weights are $k - 2 = 1$, $k - 1 = 2$, and $k = 3$. For example, $b \rightarrow_3 g$ in $G$ and thus we have the directed edge $(b, g)$ with $\omega_I((b, g)) = 3$ as shown in Figure 2. We will further explain how we use the $k$-reach graph to process a $k$-hop reachability query in Example 2 in Section 4.2. □

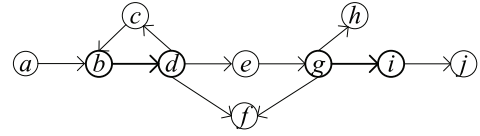

**Figure 1: An example graph $G$ (the vertex cover is $\{b, d, g, i\}$)**

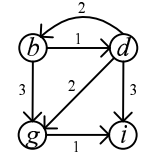

**Figure 2: The $k$-reach graph ($k = 3$), $I = (V_I, E_I, \omega_I)$, of $G$ in Figure 1**

### 4.1.3 Complexity of Constructing k-Reach

Computing the 2-approximate minimum vertex cover $S$ requires $O(m + n)$ time. Constructing the weighted, directed graph $I = (V_I, E_I, \omega_I)$ takes $O(\sum_{u \in S} |G_k(u)|)$ time, where $G_k(u)$ is the subgraph of $G$ that can be reached from $u$ in $k$ hops. Note that it is straightforward to parallelize this process if more machines or CPU cores are available.

The size of the index, i.e., the size of the graph $I$, depends on both the size of $S$ and $S_k(u)$ for each $u \in S$. However, it is difficult to derive a theoretical bound of $S$ or $S_k(u)$ for real-world graphs since they often vary significantly in characteristics with respect to $S$ and $S_k(u)$, even though some graphs may share some similar properties such as sparsity and power-law degree distribution. We are not aware of any existing work that gives a theoretical bound on the size of the minimum vertex cover for real-world graphs. Thus, we examine the size of the index experimentally for a wide range of real-world graphs.

Finally, constructing the $k$-reach index uses $O(m + n)$ memory space. Note that the constructed index is then stored on disk.

## 4.2 K-Reach: Query Processing

We now discuss how we process a $k$-hop reachability query using the $k$-reach index.

**Algorithm 2** *Query processing using k-reach*

**Input**: a directed graph, $G = (V, E)$,
       the $k$-reach index, $I = (V_I, E_I, \omega_I)$, of $G$,
       and two query vertices, $s$ and $t$
**Output**: a boolean indicator whether $s \rightarrow_k t$

/∗ *Case 1: both s and t are in the vertex cover* ∗/
1. **if**($s \in V_I$ and $t \in V_I$)
2.     **if**$((s, t) \in E_I)$
3.         **return** true;
4.     **else**
5.         **return** false;

/∗ *Case 2: only s is in the vertex cover* ∗/
6. **else if**($s \in V_I$ and $t \notin V_I$)
7.     **if**($\exists v \in inNei(t, G)$ such that
        $(s, v) \in E_I$ and $\omega_I((s, v)) \leq (k - 1)$)
8.         **return** true;
9.     **else**
10.         **return** false;

/∗ *Case 3: only t is in the vertex cover* ∗/
11. **else if**($s \notin V_I$ and $t \in V_I$)
12.     **if**($\exists v \in outNei(s, G)$ such that
        $(v, t) \in E_I$ and $\omega_I((v, t)) \leq (k - 1)$)
13.         **return** true;
14.     **else**
15.         **return** false;

/∗ *Case 4: both s and t are not in the vertex cover* ∗/
16. **else if**($s \notin V_I$ and $t \notin V_I$)
17.     **if**($\exists u \in outNei(s, G)$ and $\exists v \in inNei(t, G)$ such that
        $(u, v) \in E_I$ and $\omega_I((u, v)) \leq (k - 2)$)
18.         **return** true;
19.     **else**
20.         **return** false;

### 4.2.1 Query Processing using k-Reach

We give the algorithm for query processing using $k$-reach in Algorithm 2.

Given two query vertices, $s$ and $t$, Algorithm 2 processes the $k$-hop reachability query by considering four cases. The following theorem proves the correctness of the algorithm for processing a $k$-hop reachability query using the $k$-reach index. The proof also explains how a query is processed.

THEOREM 1. *Given a directed graph, $G = (V, E)$, the $k$-reach index, $I = (V_I, E_I, \omega_I)$, of $G$, and two query vertices, $s$ and $t$, Algorithm 2 returns* true *if $s \rightarrow_k t$ in $G$ and* false *otherwise.*

PROOF. Note that $V_I$ is a vertex cover of $G$. There are only four possible cases in processing a $k$-hop reachability query by considering the membership of $s$ and $t$ in $V_I$. Algorithm 2 processes the query according to which case the query belongs to as follows.

Case 1: both $s$ and $t$ are in $V_I$. In this case, if $s \rightarrow_k t$ in $G$, then the edge $(s, t)$ must exist in $I$. Thus, the answer to the query by Algorithm 2 is trivially correct.

Case 2: only $s$ is in $V_I$. In this case, all in-neighbors (if any) of $t$ must be in $V_I$. Otherwise if $\exists v \in inNei(t, G)$ such that $v$ is not in $V_I$, then the edge $(v, t)$ is not covered since both $v$ and $t$ are not in the vertex cover $V_I$. Thus, if $s \rightarrow_k t$ in $G$, then there must exist an in-neighbor $v$ of $t$ such that $\omega_I((s, v)) \leq (k - 1)$, since the (directed) path from $s$ to $t$ must pass through at least one in-neighbor of $t$. Therefore, it is sufficient to check whether $(s, v) \in E_I$ and $\omega_I((s, v)) \leq (k-1)$ in order to determine whether $s \rightarrow_k t$.

Case 3: only $t$ is in $V_I$. This case is similar to Case 2. Now since $s$ is not in the vertex cover $V_I$, all out-neighbors (if any) of $s$ must be in $V_I$; otherwise the edge $(s, v)$ is not covered for some $v \in outNei(s, G)$ and $v \notin V_I$. Thus, similar to Case 2, it is sufficient to check whether $(v, t) \in E_I$ and $\omega_I((v, t)) \leq (k - 1)$ in order to determine whether $s \rightarrow_k t$.

Case 4: both $s$ and $t$ are not in $V_I$. In this case, all out-neighbors (if any) of $s$ and all in-neighbors (if any) of $t$ must be in $V_I$; otherwise the edges $(s, u)$ and $(v, t)$ are not covered for some $u \in outNei(s, G), v \in inNei(t, G)$, and $u, v \notin V_I$. Thus, if $s \rightarrow_k t$ in $G$, then there must exist an out-neighbor $u$ of $s$ and an in-neighbor $v$ of $t$ such that $\omega_I((u, v)) \leq (k - 2)$, since the (directed) path from $s$ to $t$ must first go from $s$ to some $u \in outNei(s, G)$, and finally pass through some $v \in inNei(t, G)$ to $t$. Therefore, it is sufficient to check whether $(u, v) \in E_I$ and $\omega_I((u, v)) \leq (k - 2)$ in order to determine whether $s \rightarrow_k t$. □

We give an example of using the $k$-reach index to process $k$-hop reachability queries as follows. We use $s \not\rightarrow_k t$ to indicate that $t$ is **not $k$-hop reachable** from $s$.

*Example 2.* Given the graph $G$ in Figure 1 and the $k$-reach graph $I = (V_I, E_I, \omega_I)$ of $G$ in Figure 2, where $k = 3$. We discuss how we use $k$-reach to process each of four cases in Algorithm 2 as follows.

Case 1: both $s$ and $t$ are in $V_I$. Let $s = b \in V_I$. We first consider $t = g \in V_I$. Since $(b, g) \in E_I$, we have $b \rightarrow_k g$. However, if $t = i \in V_I$, then $b \not\rightarrow_k i$ since $(b, i) \notin E_I$, although $b$ can reach $i$ in $G$ (but in $4 > k = 3$ hops).

Case 2: only $s$ is in $V_I$. Let $s = d \in V_I$. If $t = h \notin V_I$, then we have $d \rightarrow_k h$ since there is an in-neighbor $g$ of $h$ such that $(d, g) \in E_I$ with $\omega_I((d, g)) = 2 \leq (k - 1) = 2$. But if $t = j \notin V_I$, then $d \not\rightarrow_k j$ since for the only in-neighbor $i$ of $j$, although $(d, i) \in E_I$, we have $\omega_I((d, i)) = 3 > (k - 1)$. We can easily verify in $G$ that $j$ is reachable from $d$ in at least 4 hops and thus not 3-hop reachable from $d$.

Case 3: only $t$ is in $V_I$. Let $s = a \notin V_I$. If $t = d \in V_I$, we have $a \rightarrow_k d$ since there is an out-neighbor $b$ of $a$ such that $(b, d) \in E_I$ with $\omega_I((b, d)) = 1 \leq (k - 1) = 2$. But if $t = g \in V_I$, then $a \not\rightarrow_k g$ since $\omega_I((b, g)) = 3 > (k - 1)$. We can easily verify in $G$ that $g$ is reachable from $a$ in at least 4 hops and thus not 3-hop reachable from $a$.

Case 4: both $s$ and $t$ are not in $V_I$. Let $s = c \notin V_I$. If $t = f \notin V_I$, we have $c \rightarrow_k f$ since there is an out-neighbor $b$ of $c$ and an in-neighbor $d$ of $f$ such that $(b, d) \in E_I$ with $\omega_I((b, d)) = 1 \leq (k - 2) = 1$. But if $t = h \notin V_I$, then $c \not\rightarrow_k h$ since $h$ has only one in-neighbor $g$ but $\omega_I((b, g)) = 3 > (k - 2)$. We can easily verify in $G$ that $h$ is reachable from $c$ in at least 5 hops and thus not 3-hop reachable from $c$. □

### 4.2.2 Complexity of Query Processing using k-Reach

The membership tests whether $s$ and $t$ belong to $V_I$ take $O(1)$ time. Checking whether an edge $(u, v)$ exists in $E_I$ and retrieving its weight take $O(\log outDeg(u, I))$ or $O(\log inDeg(v, I))$ CPU time if $I$ is stored as adjacency lists. Thus, Case 1 of Algorithm 2 takes $O(\log outDeg(s, I))$ time, Case 2 takes $O(outDeg(s, I) + inDeg(t, G))$ time, Case 3 takes $O(outDeg(s, G) + inDeg(t, I))$ time, Case 4 uses $O(\sum_{u \in outNei(s, G)} (outDeg(u, I) + inDeg(t, G)))$ time. Note that for Cases 2 to 4 we can perform intersection of the involved adjacency lists and terminate earlier as soon as an edge is found to give a true answer.

In addition to the CPU cost, to retrieve an adjacency list $L$ from disk, it also requires $O(|L|/B)$ I/Os, where $B$ is the disk block size. In practice, we have $|L| < B$ for the majority of vertices and therefore the I/O cost is small in most cases.

## 4.3 The Curse of High-Degree Vertices

According to the complexity analysis of query processing in Section 4.2.2, the query performance depends largely on the degree of a vertex in $G$ and in $I$. Many large real-world graphs have a power-law degree distribution and hence a small number of vertices may have a very high degree. For example, the singer-songwriter Lady Gaga has 40,000,000 fans on Facebook. Therefore, it is crucial to avoid having these high-degree vertices as query vertices that fall into Case 4, or even Cases 2 and 3, of Algorithm 2. Nevertheless, statistically these high-degree vertices may indeed have a higher probability to be picked as query vertices since they usually represent objects that attract more attention.

To enable these high-degree query vertices to be processed efficiently, we modify the algorithm for computing the 2-approximate minimum vertex cover in Section 4.1.1 as follows. In picking an edge in order to include its two end vertices in the vertex cover, we give higher priority to edges with either or both end vertices that have a high degree. Since most real-world graphs have only a very small percentage of high-degree vertices [25], we can easily include all such vertices in the vertex cover without sacrificing the approximation ratio. In fact, including the high-degree vertices in the vertex cover is a greedy strategy that tends to reduce the size of the vertex cover in practice, since a high-degree vertex covers more edges than a low-degree one.

Prior study has shown that for a typical real-world graph with power-law degree distribution, if the graph has 1 million vertices, then the "$h$-index" of the graph is only about 300 [10, 11]; that is, the 1 million-vertex graph contains only about $h = 300$ vertices with degree at least $h = 300$.

The vertices that have a high degree in $G$, however, also tend to have a high degree in $I$. This not only reflects a tradeoff in query performance but also increases the index size. However, this problem can be alleviated as follows. Since there are only three types of edge weight, i.e., $k$, $(k-1)$, and $(k-2)$, in $I$, we only need to use 2 bits to represent each edge weight. Thus, the set of neighbors of those high-degree vertices in $I$ can be effectively represented in a more compact way, such as *interval lists* or *partitioned word aligned hybrid compression* [28], which have been shown effective for reducing the storage size of the edge transitive closure graph for processing reachability queries. Note that with the compact representations, we only need to locate the corresponding interval or bits for query processing [28], instead of searching the list of neighbors.

## 4.4 A General $k$

We next consider if one wants to ask $k$-hop reachability queries for different values of $k$. In this case, a specific $k$-reach index (i.e., the index is built on a specific value of $k$) is not able to handle a general $k$. However, we note that if the index can process $k$-hop reachability queries with a general $k$, then the index is essentially an index for shortest-path distance queries. To the best of our knowledge, so far there is no efficient index for answering shortest-path or distance queries in directed general graphs (see Section 3.5). We discuss two possible approaches of handling a general $k$ as follows.

First, our index can be easily generalized to process distance queries by keeping the distance information between any two vertices in the vertex cover, which can be computed by doing a full BFS instead of a $k$-hop BFS in Line 5 of Algorithm 1. This requires $\lg d$ bits for each edge weight (instead of 2 bits as with a specific $k$), where $d$ is the diameter of the input graph. However, in order to answer distance queries, the $k$-reach graph for a general $k$ is a complete graph, while the $k$-reach graph for a specific $k$ is a sparse graph. Thus, this generalization of $k$-reach works only for small $d$ and small graphs.

The second approach is based on the observation that in many applications (e.g., message broadcasting in sensor networks and connectivity in social networks), the influence/significance of a $k$-hop neighbor on/to a vertex decreases quickly as $k$ increases. Thus, we can build $(\lg d)$ $i$-reach indexes, where $i = 2^1, \cdots, 2^{\lg d} = d$. Then, to answer whether $s$ can reach $t$ within $k$ hops, we use the $2^{\lceil \lg k \rceil}$-reach index. If $s$ can reach $t$ within $k$ hops, or $s$ cannot reach $t$ within $2^{\lceil \lg k \rceil}$ hops, then clearly the index gives a correct answer. Otherwise, the index gives an approximate answer that $s$ can reach $t$ within $k'$ hops, where $k < k' \le 2^{\lceil \lg k \rceil}$ hops. Thus, this method gives higher significance to smaller $k$ because the range of $k'$ becomes larger as $k$ becomes larger.

The overall space taken by the $(\lg d)$ $i$-reach indexes is approximately $\lg d$ times the space of a single $k$-reach, as the size of the $i$-reach index remains rather stable for different values of $i$ (as tested in Section 6.3). For many real-world graphs, $d$ is relatively small according to the six-degree-of-separation theory. For the 15 real datasets used in our experiments, $d$ is around 10 for most of them and the largest $d$ is 24 (see Table 2).

If accuracy is critical for some applications, one may even build the $i$-reach indexes for each $i = 2, \cdots, d$, to obtain exact answer for $k$-hop reachability queries of any $k$. Given the small value of $d$ in many real-world graphs, the storage size may be affordable. Or some applications may be only interested in small values of $k$ and hence construct only the $k$-reach index for the first few $k$ (e.g., for $k < 6$). Note that none of the existing reachability indexes support $k$-hop reachability even if we limit $k$ to a small number.

## 5. A TRADEOFF BETWEEN INDEXING AND QUERYING

It is well-known that there is a tradeoff between the cost of constructing a reachability index and the cost of processing reachability queries [33]. More specifically, one can spend $O(nm)$ time to construct an index of size $O(n^2)$ so that any query can be answer in $O(1)$ time. On the contrary, one can answer a query in $O(n + m)$ time without an index. All the existing reachability indexes lie somewhere between these two extremes, attempting to attain a low querying cost with a reasonable indexing cost.

The tradeoff between indexing cost and querying cost also exists in $k$-reach. This section proposes a method to reduce the index size when it becomes large.

## 5.1 An h-Hop VC-based k-Reach Index

The idea to reduce the indexing time and index storage size is based on the concept of a *path-based vertex cover*. We first present the concept of the path-based vertex cover and then discuss how to construct a new $k$-reach index.

### 5.1.1 Minimum h-Hop Vertex Cover

We first define the notion of **h-hop vertex cover** [4] as follows.
A set of vertices, $S$, is an $h$-hop vertex cover of a graph $G = (V, E)$, where $1 \le h < n$, if for every path $P = \langle v_0, \cdots, v_h \rangle$ of length $h$ in $G$, we have $(\{v_0, \cdots, v_h\} \cap S) \ne \emptyset$. An $h$-hop vertex cover $S$ is called a *minimum $h$-hop vertex cover* of $G$ if $S$ has the smallest size among all $h$-hop vertex covers of $G$.

Trivially, $V$ itself is an $h$-hop vertex cover of $G$ for any valid $h$. In fact, a vertex cover of $G$ is also an $h$-hop vertex cover of $G$ for any valid $h$. More specifically, we have the following lemma.

LEMMA 1. *An $i$-hop vertex cover of a graph $G$ is also a $j$-hop vertex cover of $G$, where $1 \leq i \leq j < n$.*

PROOF. Since every path of length $i$ in $G$ is covered by some vertex in the $i$-hop vertex cover, every path $P$ of length $j$ is clearly covered by the same vertex that covers the length-$i$ sub-path of $P$. Thus, an $i$-hop vertex cover of $G$ is also a $j$-hop vertex cover of $G$. □

The following corollary follows from Lemma 1.

COROLLARY 1. *Let $S_i$ be a minimum $i$-hop vertex cover of a graph $G$ and $S_j$ be a minimum $j$-hop vertex cover of $G$, where $1 \leq i \leq j < n$. Then, $|S_j| \leq |S_i|$.*

PROOF. The proof is trivial since by Lemma 1, one can use $S_i$ as the minimum $j$-hop vertex cover if $|S_j| > |S_i|$, contradicting that $S_j$ is minimum. □

Corollary 1 shows that we can construct an index based on a minimum $h$-hop vertex cover of a larger $h$ in order to obtain a small index size.

We can easily prove that the general problem of computing a minimum $h$-hop vertex cover is also NP-hard, since the special case minimum 1-hop vertex cover, or simply minimum vertex cover, is NP-hard. We give a polynomial time algorithm that computes a $(h+1)$-*approximate minimum $h$-hop vertex cover* as follows.

We randomly select a path $P = \langle v_0, \cdots, v_h \rangle$ of length $h$ in $G$, and add all vertices on the path, i.e., $\{v_0, \cdots, v_h\}$, to $S$. Then, we remove all vertices in $\{v_0, \cdots, v_h\}$ from $G$, together with all edges incident on these vertices. Note that all edges incident on $\{v_0, \cdots, v_h\}$, whether in-edges or out-edges, can be removed from $G$ because any length-$h$ path containing any of these edges is covered by some vertices in $\{v_0, \cdots, v_h\}$. This process is repeated until no path of length $h$ exists in $G$.

It is not difficult to see that the above algorithm is $(h+1)$-approximate. Let $C$ be a minimum $h$-hop vertex cover of $G$. For every path $P = \langle v_0, \cdots, v_h \rangle$ selected, we include $\{v_0, \cdots, v_h\}$ in $S$ in the above process. Note that at least one vertex in $\{v_0, \cdots, v_h\}$ must be in $C$, because otherwise $P$ is not covered by any vertex in $C$. Thus, we have $|S| \leq (h+1)|C|$.

Bresar et al. [4] studied the theoretical upper bound on the size of a minimum $h$-hop vertex cover of general graphs, outerplanar graphs, and trees. For general graphs, they show that the size of a minimum $h$-hop vertex cover is at most $(n - \frac{h-1}{h} \sum_{v \in V} \frac{2}{1+Deg(v,G)})$. This upper bound is for every general graph and therefore often too loose. For most real-world graphs, the size of a minimum $h$-hop vertex cover is reasonably small for our indexing purpose even when a small $h$ is used.

We also remark that, by Lemma 1 and Corollary 1, for index construction we always have a smaller $(j+1)$-approximate minimum $j$-hop vertex cover than an $(i+1)$-approximate minimum $i$-hop vertex cover, where $1 \leq i \leq j < n$. This is true because, if any $(i+1)$-approximate minimum $i$-hop vertex cover is smaller, we can always simply use it as a $(j+1)$-approximate minimum $j$-hop vertex cover.

### 5.1.2 The (h,k)-Reach Index

The structure of the new $h$-hop vertex cover based $k$-reach index is given as follows.

*Definition 2* (($h,k$)-REACH). *Given a directed graph $G = (V, E)$, an $h$-hop vertex cover $S$ of $G$, and an integer $k$, where $h < k/2$, the **$h$-hop vertex cover based $k$-reach**, or **($h,k$)-reach** in short, of $G$ is a weighted, directed graph $H = (V_H, E_H, \omega_H)$ defined as follows.*

- $V_H = S$.
- $E_H = \{(u,v) : u, v \in S, u \to_k v\}$.
- $\omega_H$ is a function that assigns a weight to each edge $e = (u,v) \in E_H$ as follows: $\omega_H(e) = (k-i)$ iff $u \to_{k-i} v$ and $u \not\to_{k-i-1} v$, where $0 \leq i \leq 2h$, and $u \not\to_{k-i-1} v$ means that $v$ cannot be reached from $u$ within $(k-i-1)$ hops.

Note that the value of $h$ is set as $h < k/2$ because we consider at most $k/2$ hops from a neighbor of $s$ and at most $k/2$ hops to a neighbor of $t$ when both query vertices $s$ and $t$ are not in the vertex cover.

The index construction process is identical to Algorithm 1, except that we use an $(h+1)$-approximate minimum $h$-hop vertex cover instead of a 2-approximate minimum vertex, and assign the edge weight according to Definition 2.

The complexity analysis of Algorithm 1 applies in the same way to the complexity of index construction for $(h,k)$-reach. However, the indexing time and the index storage space are both reduced because, according to the discussion in Section 5.1.1, the size of an $(h+1)$-approximate minimum $h$-hop vertex cover is always smaller than that of a 2-approximate minimum vertex.

We give an example of an $(h,k)$-reach index as follows.

*Example 3.* Given the graph $G$ in Figure 3, which is the same graph of Figure 1 in Example 1 but this time we apply the $h$-hop vertex cover of $G$. Let $h = 2$. Assume that the 3-approximate algorithm randomly picks the 2-hop path, $\langle d, e, g \rangle$, in $G$. Then, $\{d, e, g\}$ forms a 3-approximate minimum 2-hop vertex cover of $G$. We can easily verify that $\{d, e, g\}$ is indeed a 2-hop vertex cover of $G$, since every path of length 2 in $G$ is incident on at least one of the vertices in $\{d, e, g\}$.

Let $k = 5$. The $(2,5)$-reach graph, $H = (V_H, E_H, \omega_H)$, of $G$ is shown in Figure 4. We will further illustrate how we process a $k$-hop reachability query using $H$ in Example 4 in Section 5.2. □

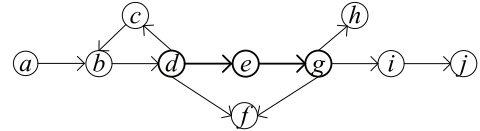

**Figure 3: An example graph $G$ (the 2-hop vertex cover is $\{d, e, g\}$)**

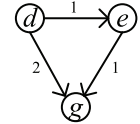

**Figure 4: The $(h,k)$-reach graph ($h = 2, k = 5$), $H = (V_H, E_H, \omega_H)$, of $G$ in Figure 3**

## 5.2 Query Processing using (h,k)-Reach

We now discuss how we use $(h,k)$-reach to process a $k$-hop reachability query, as shown in Algorithm 3. Before we discuss the details of the algorithm, we need to first define a few notations that are used in the algorithm description and complexity analysis.

We denote the set of *$i$-hop in-neighbors* of a vertex $v$ in $G$ by $inNei_i(v, G) = \{u : u \to_i v\}$, and the *$i$-hop in-degree* of $v$ in

**Algorithm 3** *Query processing using (h,k)-reach*

**Input**: a directed graph, $G = (V, E)$,
   the $(h, k)$-reach index,
   $H = (V_H, E_H, \omega_H)$, of $G$,
   and two query vertices, $s$ and $t$

**Output**: a boolean indicator whether $s \rightarrow_k t$

/∗ *Case 1: both s and t are in the h-hop vertex cover* ∗/
1.  **if**($s \in V_H$ and $t \in V_H$)
2.     **if**($(s, t) \in E_H$)
3.        **return** `true`;
4.     **else**
5.        **return** `false`;

/∗ *Case 2: only s is in the h-hop vertex cover* ∗/
6.  **else if**($s \in V_H$ and $t \notin V_H$)
7.     **if**($\exists v \in inNei_i(t, G)$, where $1 \leq i \leq h$,
      such that $(s, v) \in E_H$ and $\omega_H((s, v)) \leq (k - i)$)
8.        **return** `true`;
9.     **else**
10.       **return** `false`;

/∗ *Case 3: only t is in the h-hop vertex cover* ∗/
11. **else if**($s \notin V_H$ and $t \in V_H$)
12.    **if**($\exists v \in outNei_i(s, G)$, where $1 \leq i \leq h$,
      such that $(v, t) \in E_H$ and $\omega_H((v, t)) \leq (k - i)$)
13.       **return** `true`;
14.    **else**
15.       **return** `false`;

/∗ *Case 4: both s and t are not in the h-hop vertex cover* ∗/
16. **else if**($s \notin V_H$ and $t \notin V_H$)
17.    **if**($\exists u \in outNei_i(s, G)$ and $\exists v \in inNei_j(t, G)$,
      where $1 \leq i \leq h$ and $1 \leq j \leq h$,
      such that $(u, v) \in E_H$ and $\omega_H((u, v)) \leq (k - i - j)$)
18.       **return** `true`;
19.    **else**
20.       **return** `false`;

$G$ as $inDeg_i(v, G) = |inNei_i(v, G)|$. Similarly, we denote the set of $i$-**hop out-neighbors** of $v$ in $G$ by $outNei_i(v, G) = \{u : v \rightarrow_i u\}$, and the $i$-**hop out-degree** of $v$ in $G$ as $outDeg_i(v, G) = |outNei_i(v, G)|$.

Algorithm 3 is similar to Algorithm 2 except that it considers the $i$-hop neighbors of a query vertex $v$ if $v$ is not in the $h$-hop vertex cover, while Algorithm 2 considers the direct neighbors, i.e., the 1-hop neighbors of $v$ only.

The following theorem states the correctness of the algorithm.

THEOREM 2. *Given a directed graph, $G = (V, E)$, the $(h, k)$-reach index, $H = (V_H, E_H, \omega_H)$, of $G$, and two query vertices, $s$ and $t$, Algorithm 3 returns* `true` *if $s \rightarrow_k t$ in $G$ and* `false` *otherwise.*

PROOF. The proof follows a similar logic as that of Theorem 1. □

We give an example of query processing using the $(h, k)$-reach index as follows.

*Example 4.* Given the graph $G$ in Figure 3 and the $(h, k)$-reach graph $H = (V_H, E_H, \omega_H)$ of $G$ in Figure 4, where $h = 2$ and $k = 5$. We discuss how we use $H$ to process each of four cases in Algorithm 3 as follows.

Case 1: both $s$ and $t$ are in $V_H$. Let $s = e \in V_H$. If $t = g \in V_H$, then $e \rightarrow_k g$ since $(e, g) \in E_H$. On the contrary, if $t = d \in V_H$, then $e \nrightarrow_k d$ since $(e, d) \notin E_H$.

Case 2: only $s$ is in $V_H$. Let $s = d \in V_H$. Consider $t = h \notin V_H$. Since $g \in inNei_1(h, G)$ such that $(d, g) \in E_H$ and $\omega_H((d, g)) = 2 \leq (k - 1) = 4$, we have $d \rightarrow_k h$. Note that $e \in inNei_2(h, G)$ can also be used to process the query but the process terminates as soon as a vertex in $V_H$ can be used to answer the query. Now consider $t = a \notin V_H$. Since $inNei_i(a, G) = \emptyset$, where $1 \leq i \leq h = 2$, we have $d \nrightarrow_k a$.

Case 3: only $t$ is in $V_H$. Let $s = a \notin V_H$. If $t = g \in V_H$, we have $a \rightarrow_k g$ since we have $d \in outNei_2(a, G)$ such that $(d, g) \in E_H$ and $\omega_H((d, g)) = 2 \leq (k - 2) = 3$.

Case 4: both $s$ and $t$ are not in $V_H$. Let $s = a \notin V_H$. If $t = i \notin V_H$, we have $a \rightarrow_k i$ since $d \in outNei_2(a, G)$ and $g \in inNei_1(i, G)$ such that $(d, g) \in E_H$ and $\omega_H((d, g)) = 2 \leq (k - 2 - 1) = 2$. However, if $t = j \notin V_H$, then $a \nrightarrow_k j$ since we only have $d \in outNei_2(a, G)$ and $g \in inNei_2(j, G)$, but $\omega_H((d, g)) = 2 > (k - 2 - 2) = 1$. We can easily verify in $G$ that $a$ can reach $i$ in 5 hops but can reach $j$ in at least $6 > k = 5$ hops. □

Compared with the $k$-reach index, the $(h, k)$-reach index trades querying time for indexing time and index storage size. The increased time complexity for query processing is mainly due to that the search space is expanded from the direct neighbors to the $i$-hop neighbors, where $1 \leq i \leq h$. We analyze the querying complexity as follows.

In the worst case, query processing for Case 1 of Algorithm 3 takes $O(\log outDeg(s, H))$ time, Case 2 takes $O(outDeg(s, H) + inDeg_i(t, G))$ time, Case 3 takes $O(outDeg_i(s, G) + inDeg(t, H))$ time, and Case 4 takes $O(\sum_{u \in outNei_i(s,G)}(outDeg(u, H) + inDeg_j(t, G)))$ time, where $1 \leq i \leq h$ and $1 \leq j \leq h$.

In most cases, query processing does not search all the $i$-hop or $j$-hop in/out-neighbors, but terminates earlier as soon as a match is found.

## 6. EXPERIMENTAL EVALUATION

We evaluate the performance of $k$-reach, by comparing with the state-of-the-art indexes for processing classic reachability queries, including *path-tree cover*[1] (**PTree**) [24], **3-hop**[2] [23], **GRAIL** [32], and *Partitioned Word Aligned Hybrid compression* (**PWAH**) [28].

All systems, both ours and others we compared with, were implemented in C++ and compiled using the same gcc compiler. We ran all the experiments on a machine with an Intel Quad Q9400 2.66GHz CPU and 4GB RAM, running Scientific Linux release 6.0. The experiments were run for 10 times and the results were found to be consistent over the 10 runs.

### 6.1 Datasets

We conducted our experiments on a list of 15 real datasets that are popularly used to assess the performance of graph reachability indexes in the existing works [23, 24, 28, 32].

The datasets, `AgroCyc`, `Anthra`, `Ecoo`, `Human`, `Mtbrv`, and `Vchocyc`, are from EcoCyc (*ecocyc.org*) and describe the genome

---

[1] We only recently learnt that the PTree code we downloaded from the authors' website has been updated. Some comparison results between their updated code and other reachability indexes can be found in [22].

[2] Note that *3-hop* is only the name of the index [23] for processing classic reachability queries, and does not imply 3-hop reachability.

and biochemical machinery of E. coli K-12 MG1655. The datasets, aMaze and Kegg, are metabolic networks [27]. The datasets, Nasa and Xmark, are XML documents [24]. The datasets, ArXiv (*arxiv.org*), CiteSeer (*citeseer.ist.psu.edu*), and PubMed (*pubmedcentral.nih.gov*), are citation networks. The GO dataset (*www.geneontology.org*) is a gene ontology graph. The YAGO dataset (*mpi-inf.mpg.de/yago-naga/yago*) is a graph that describes the structure of relationships among terms in the semantic knowledge database YAGO.

Table 2 shows the number of vertices and edges ($|V|$ and $|E|$), the maximum vertex degree ($Deg_{max}$), the diameter ($d$), and the median length of all shortest paths ($\mu$) of the datasets. We also show the number of vertices and edges of the corresponding DAG of each dataset ($|V_{DAG}|$ and $|E_{DAG}|$), since the existing indexes we compare with all assume that the input graph is a DAG.

**Table 2: Datasets**

|          | $|V|$  | $|E|$  | $|V_{DAG}|$ | $|E_{DAG}|$ | $Deg_{max}$ | $d$ | $\mu$ |
|----------|--------|--------|-------------|-------------|-------------|-----|-------|
| AgroCyc  | 13,969 | 17,694 | 12,684      | 13,657      | 5,488       | 10  | 2     |
| aMaze    | 11,877 | 28,700 | 3,710       | 3,947       | 3,097       | 11  | 2     |
| Anthra   | 13,766 | 17,307 | 12,499      | 13,327      | 5,401       | 10  | 2     |
| ArXiv    | 6,000  | 66,707 | 6,000       | 66,707      | 700         | 20  | 4     |
| CiteSeer | 10,720 | 44,258 | 10,720      | 44,258      | 192         | 18  | 3     |
| Ecoo     | 13,800 | 17,308 | 12,620      | 13,575      | 5,435       | 10  | 2     |
| GO       | 6,793  | 13,361 | 6,793       | 13,361      | 71          | 11  | 3     |
| Human    | 40,051 | 43,879 | 38,811      | 39,816      | 28,571      | 10  | 2     |
| Kegg     | 14,271 | 35,170 | 3,617       | 4,395       | 3,282       | 16  | 2     |
| Mtbrv    | 10,697 | 13,922 | 9,602       | 10,438      | 4,005       | 12  | 2     |
| Nasa     | 5,704  | 7,942  | 5,605       | 6,538       | 32          | 22  | 7     |
| PubMed   | 9,000  | 40,028 | 9,000       | 40,028      | 432         | 11  | 4     |
| Vchocyc  | 10,694 | 14,207 | 9,491       | 10,345      | 3,917       | 10  | 2     |
| Xmark    | 6,483  | 7,654  | 6,080       | 7,051       | 887         | 24  | 5     |
| YAGO     | 6,642  | 42,392 | 6,642       | 42,392      | 2,371       | 9   | 1     |

## 6.2 Performance Comparison with the Existing Indexes

Since there is no other existing work for processing $k$-hop reachability queries, we compare with the state-of-the-art indexes for processing classic reachability queries. Such a performance comparison is not meaningless because a reachability query is a special case of a $k$-hop reachability query, that is, when $k = n$ or $k = \infty$. However, we remark that processing $k$-hop reachability queries has a higher cost than processing reachability queries since some distance information needs to be handled for $k$-hop reachability.

We compare with PTree [24], 3-hop [23], GRAIL [32] and PWAH [28], on the performance of both index construction performance and query processing. We denote our index as ***n*-reach**, for the case when $k = n$, which is essentially an index for processing reachability queries.

### 6.2.1 Performance of Index Construction

Table 3 reports the index construction time of all the indexes for all the datasets. The result shows that the construction of the $n$-reach index is faster than that of the PTree index in all cases. On average, constructing $n$-reach is approximately 7.9 times faster than constructing PTree. Compared with the construction of GRAIL and PWAH, however, the construction of $n$-reach is slower, though the time difference is rather small compared with the difference between our indexing time and that of PTree. More importantly, we will show next in Table 5 that query processing by $n$-reach is significantly faster than both GRAIL and PWAH. For the construction of the 3-hop index, we were not able to obtain the result for most of the datasets (as indicated by "-" in Tables 3 to 5) due to long running time or large memory consumption. We note that the 3-hop index was primarily designed for processing denser graphs.

**Table 3: Index construction time (elapsed time in milliseconds) of *n*-reach, PTree, 3-hop, GRAIL, and PWAH (shortest time shown in bold)**

|          | *n*-reach | PTree    | 3-hop   | GRAIL  | PWAH   |
|----------|-----------|----------|---------|--------|--------|
| AgroCyc  | 27.71     | 129.14   | -       | 10.86  | **4.40** |
| aMaze    | 18.09     | 476.69   | 959,821 | **2.92** | 7.01 |
| Anthra   | 24.08     | 123.43   | -       | 10.74  | **3.90** |
| ArXiv    | 352.51    | 6,319.66 | -       | **10.58** | 111.00 |
| CiteSeer | 245.46    | 403.35   | 44,328  | **16.04** | 93.26 |
| Ecoo     | 26.70     | 129.74   | -       | 10.88  | **4.47** |
| GO       | 106.84    | 110.83   | 11,914  | **6.50** | 19.57 |
| Human    | 67.78     | 397.05   | -       | 41.45  | **6.71** |
| Kegg     | 21.01     | 537.17   | -       | **2.92** | 6.77 |
| Mtbrv    | 20.24     | 98.13    | -       | 7.92   | **3.86** |
| Nasa     | 57.93     | 62.22    | 13,739  | **4.51** | 10.54 |
| PubMed   | 166.23    | 437.16   | 73,243  | **11.63** | 70.63 |
| Vchocyc  | 19.77     | 97.34    | -       | 7.60   | **4.00** |
| Xmark    | 44.50     | 136.87   | 68,219  | **4.96** | 11.53 |
| YAGO     | 32.47     | 282.45   | 5,006   | **9.47** | 36.49 |

**Table 4: Index size (in MB) of *n*-reach, PTree, 3-hop, GRAIL, and PWAH (smallest size shown in bold)**

|          | *n*-reach | PTree | 3-hop | GRAIL | PWAH |
|----------|-----------|-------|-------|-------|------|
| AgroCyc  | 0.39      | 0.29  | -     | **0.19** | 0.44 |
| aMaze    | 0.13      | 0.09  | 5.41  | **0.06** | 0.22 |
| Anthra   | 0.36      | 0.29  | -     | **0.19** | 0.42 |
| ArXiv    | 1.61      | 0.38  | -     | **0.09** | 2.46 |
| CiteSeer | 3.17      | 0.45  | 0.20  | **0.16** | 3.08 |
| Ecoo     | 0.40      | 0.29  | -     | **0.19** | 0.43 |
| GO       | 1.28      | 0.20  | 0.11  | **0.10** | 0.63 |
| Human    | 1.17      | 0.89  | -     | **0.59** | 1.25 |
| Kegg     | 0.16      | 0.08  | -     | **0.06** | 0.23 |
| Mtbrv    | 0.29      | 0.22  | -     | **0.15** | 0.34 |
| Nasa     | 0.66      | 0.13  | **0.06** | 0.09 | 0.40 |
| PubMed   | 2.03      | 0.50  | 0.29  | **0.14** | 2.80 |
| Vchocyc  | 0.28      | 0.22  | -     | **0.14** | 0.33 |
| Xmark    | 0.49      | 0.13  | 0.43  | **0.09** | 0.45 |
| YAGO     | 0.48      | 0.22  | **0.09** | 0.10 | 0.96 |

The memory consumption of constructing the $n$-reach index, as well as GRAIL and PWAH, is all very small. These three indexes use similar amount of memory, which varies from about 4 MB to 10 MB from dataset to dataset. The memory consumption of constructing the PTree and 3-hop indexes is considerably higher.

Table 4 reports the storage size (on disk) of the various indexes for the different datasets. The result shows that the size of GRAIL is the smallest among almost all datasets. The size of $n$-reach is slightly larger than that of PTree but smaller than that of PWAH for most datasets. Overall, the difference in size is small and affordable with today's disk storage size. However, we remark that our index is mainly designed for processing $k$-hop reachability queries, and therefore it is reasonable that it uses more space since more distance information is required to be indexed.

### 6.2.2 Performance of Query Processing

To compare the performance of query processing of the various indexes, we randomly generated 1 million queries. We emphasize that, as we will explain later in details in Table 8, these queries are not chosen to favor the performance of our index.

Table 5 reports the total time used to process the 1 million queries by the different indexes. The memory consumption of query processing by the various indexes is roughly the same as their respective index size.

**Table 5: Total running time (elapsed time in milliseconds) of *n*-reach, PTree, 3-hop, GRAIL, and PWAH, for processing 1 million randomly generated queries (shortest time shown in bold)**

|          | *n*-reach | PTree | 3-hop | GRAIL | PWAH |
|----------|-----------|-------|-------|-------|------|
| AgroCyc  | **5.50**  | 17.74 | -     | 135.14 | 15.68 |
| aMaze    | **14.39** | 20.68 | 28404.20 | 2982.61 | 39.71 |
| Anthra   | **5.39**  | 17.66 | -     | 121.12 | 14.92 |
| ArXiv    | 87.86     | **75.28** | - | 2032.96 | 311.55 |
| CiteSeer | 115.64    | **58.28** | 1225.25 | 268.33 | 339.23 |
| Ecoo     | **5.47**  | 17.73 | -     | 154.41 | 15.77 |
| GO       | **27.00** | 35.77 | 455.83 | 113.46 | 59.10 |
| Human    | **5.95**  | 28.48 | -     | 300.23 | 13.35 |
| Kegg     | **16.27** | 22.51 | -     | 4030.89 | 44.52 |
| Mtbrv    | **5.47**  | 17.48 | -     | 104.15 | 16.12 |
| Nasa     | **18.26** | 23.62 | 359.16 | 64.27 | 43.94 |
| PubMed   | **39.31** | 103.44 | 1198.70 | 239.40 | 368.44 |
| Vchocyc  | **5.49**  | 17.72 | -     | 103.23 | 16.13 |
| Xmark    | **14.49** | 22.02 | 491.44 | 245.11 | 69.78 |
| YAGO     | 106.25    | **42.32** | 705.09 | 116.43 | 137.09 |

The result shows that query processing by *n*-reach is significantly faster than all the other indexes. On average, *n*-reach is 2.2 times faster than PTree in query processing, but is 7.9 times faster in index construction (see Table 3). Query processing by *n*-reach is also 3.2 times faster than PWAH on average, with slightly longer indexing time. Compared with GRAIL and 3-hop, *n*-reach is a clear winner since the running time of *n*-reach is up to orders of magnitude shorter.

### 6.2.3 Overall Performance

Table 6 gives a ranking on the performance of the various indexes for both index construction and query processing. Each ranking is obtained based on the overall ranking of the performance results of the indexes reported in Tables 3, 4, and 5, respectively.

**Table 6: Performance ranking for *n*-reach, PTree, 3-hop, GRAIL, and PWAH on indexing time, index size, and query processing time (1 is the best)**

|               | *n*-reach | PTree | 3-hop | GRAIL | PWAH |
|---------------|-----------|-------|-------|-------|------|
| Indexing time | 3         | 4     | 5     | 1     | 2    |
| Index size    | 3         | 2     | 5     | 1     | 4    |
| Querying time | 1         | 2     | 5     | 4     | 3    |

For index construction, GRAIL is the clear winner. The indexing performance of *n*-reach, PTree, and PWAH is comparable overall, as they are better than the other in indexing time but worse in index size, and no one is the clear winner among them.

For query processing, *n*-reach is the clear winner, as Table 3 shows that *n*-reach is the fastest in almost all cases and on average a few times faster than PTree and PWAH. The query performance of GRAIL, however, cannot match up with its indexing performance and is up to two orders of magnitude slower than *n*-reach.

Overall, given its reasonable indexing performance, the superior query performance of *n*-reach makes it a good choice even for processing classic reachability queries, even though the index is primarily designed for processing *k*-hop reachability queries. This result is surprising since the processing of *k*-hop reachability queries needs to handle some distance information as well and hence is expected to take longer time. Therefore, the result suggests the effectiveness of our method for indexing graph reachability in general.

## 6.3 Performance of k-Reach

In this subsection, we investigate the performance of *k*-reach over a range of values of *k*. We test $k = 2, 4, 6, \mu$, and $n$, where $\mu$ is the median length of all shortest paths in the corresponding dataset (see Table 2).

In Section 6.2 we have shown that the construction of the *n*-reach index is efficient and competitive with the existing reachability indexes. The construction time and index size of *k*-reach for other values of *k* is only slightly lower. We thus omit the details and report only the results of query processing by *k*-reach.

We use the same set of randomly generated queries used in Section 6.2. Note that for the same query, $s$ and $t$, the answer whether $t$ is reachable from $s$ within $k$ hops may be different for different values of $k$.

### 6.3.1 Performance of Different k-Reach Indexes, k-Hop BFS, and Shortest-Path Distance Index

Table 7 reports the total running time of processing the 1 million queries by *k*-reach, for $k = 2, 4, 6, \mu, n$. The result shows that the performance of the different *k*-reach index is stable with the different values of *k* for all the datasets. Note that the values of *k* ranging from 2 to *n* cover the two extreme ends ($k = 1$ is trivial since it only needs to check edge existence). Thus, the result demonstrates the efficiency of the *k*-reach index for processing *k*-hop reachability queries of any *k*.

**Table 7: Total running time (elapsed time in milliseconds) of *k*-reach for $k = 2, 4, 6, \mu, n$, and $\mu$-BFS and $\mu$-dist for processing 1 million randomly generated queries**

|          | 2-reach | 4-reach | 6-reach | $\mu$-reach | *n*-reach | $\mu$-BFS | $\mu$-dist |
|----------|---------|---------|---------|-------------|-----------|-----------|------------|
| AgroCyc  | 5.47    | 5.49    | 5.47    | 5.56        | 5.50      | 6666.61   | 81.32      |
| aMaze    | 14.38   | 14.42   | 14.40   | 14.39       | 14.39     | 9145.64   | 193.71     |
| Anthra   | 5.43    | 5.36    | 5.36    | 5.33        | 5.39      | 6662.71   | 73.47      |
| ArXiv    | 90.08   | 84.64   | 87.66   | 88.84       | 87.86     | 17645.10  | 30391.09   |
| CiteSeer | 116.44  | 117.08  | 107.72  | 116.50      | 115.64    | 7016.10   | 1392.21    |
| Ecoo     | 5.48    | 5.47    | 5.50    | 5.43        | 5.47      | 6667.16   | 78.18      |
| GO       | 26.99   | 27.00   | 26.97   | 27.00       | 27.00     | 6794.95   | 673.48     |
| Human    | 5.98    | 6.02    | 6.09    | 6.03        | 5.95      | 6756.70   | 45.42      |
| Kegg     | 16.16   | 16.32   | 16.22   | 16.12       | 16.27     | 9525.80   | 206.25     |
| Mtbrv    | 5.49    | 5.48    | 5.47    | 5.46        | 5.46      | 6656.73   | 90.73      |
| Nasa     | 18.26   | 18.30   | 18.24   | 18.23       | 18.26     | 6852.91   | 554.70     |
| PubMed   | 39.25   | 39.37   | 39.52   | 39.36       | 39.31     | 7301.46   | 1079.70    |
| Vchocyc  | 5.49    | 5.48    | 5.50    | 5.46        | 5.49      | 6678.73   | 90.62      |
| Xmark    | 14.38   | 14.41   | 14.46   | 14.42       | 14.49     | 7145.60   | 132.90     |
| YAGO     | 113.01  | 106.41  | 105.85  | 101.67      | 106.25    | 6723.07   | 586.10     |

Since the *k*-reach index is the first index for processing *k*-hop reachability queries, we also show in Table 7 the time taken to process the queries by performing BFS for $k = \mu$ steps, denoted by $\mu$-**BFS**, as well as the state-of-the-art index for processing distance queries in directed graphs [13], denoted by $\mu$-**dist**. Note that the indexes [30, 31] do not work for directed graphs.

The query processing time of $\mu$-BFS and $\mu$-dist is reported here as a reference only. However, the result does suggest the need of an index for processing *k*-hop reachability queries, since $\mu$-reach is up to three orders of magnitude faster than $\mu$-BFS and up to two

1301

orders of magnitude faster than $\mu$-dist. Moreover, constructing $\mu$-dist takes 14 to 81 (average 44.25) times longer than $\mu$-reach, and uses 2.3 times more space on average.

### 6.3.2 The Four Categories of Queries

In Algorithm 2 we show that there are four cases in processing a $k$-hop reachability query. The complexity analysis in Section 4.2.2 shows that processing a Case 1 query has the lowest time complexity, and processing a Case 4 query is the most costly, while the cost of processing a Case 2 or 3 query lies in the middle. Thus, we want to examine whether the queries used in the experiments may favor the performance of $k$-reach.

Table 8 reports the percentage of queries in each of the four cases in Algorithm 2, for the 1 million queries tested. We can see that for most of the datasets, the majority of the queries are Case 4 queries, while the minority are Case 1 queries. This distribution is mainly because the size of the vertex cover is only a small percentage of the total number of vertices in the graph and hence a randomly selected vertex has a lower probability being in the vertex cover.

**Table 8: Percentage of queries (among the 1 million randomly generated queries) in each case of Algorithm 2**

|          | Case 1 | Case 2 | Case 3 | Case 4 |
|----------|--------|--------|--------|--------|
| AgroCyc  | 0.10   | 2.98   | 2.96   | 93.97  |
| aMaze    | 1.65   | 11.19  | 11.23  | 75.93  |
| Anthra   | 0.08   | 2.73   | 2.79   | 94.40  |
| ArXiv    | 41.94  | 22.79  | 22.88  | 12.38  |
| CiteSeer | 19.15  | 24.62  | 24.62  | 31.61  |
| Ecoo     | 0.10   | 3.02   | 3.05   | 93.83  |
| GO       | 19.18  | 24.63  | 24.66  | 31.53  |
| Human    | 0.01   | 0.94   | 0.96   | 98.09  |
| Kegg     | 2.92   | 14.17  | 14.21  | 68.71  |
| Mtbrv    | 0.15   | 3.66   | 3.67   | 92.52  |
| Nasa     | 10.80  | 22.12  | 22.03  | 45.05  |
| PubMed   | 15.12  | 23.77  | 23.71  | 37.40  |
| Vchocyc  | 0.15   | 3.65   | 3.68   | 92.53  |
| Xmark    | 4.06   | 16.08  | 16.10  | 63.75  |
| YAGO     | 1.55   | 10.96  | 10.89  | 76.60  |

We calculated that, on average, it takes 12 times longer to process a Case 4 query than a Case 1 query, and 5 to 6 times longer to process a Case 2 or Case 3 query than a Case 1 query. Thus, we can conclude that $k$-reach is efficient for processing any general $k$-hop reachability queries, since our experimental results were not obtained based on queries that were chosen to favor the performance of our algorithm (i.e., Case 1 queries).

## 6.4 Performance of (h,k)-Reach

In this subsection, we study the tradeoff between indexing cost and querying cost by using the $(h, k)$-reach index.

Table 9 first reports the size of the vertex cover and that of the 2-hop vertex cover (in the first two columns of the table). Considering that the size of the vertex cover is already small (cf. the sizes of the datasets in Table 2), the size of the 2-hop vertex cover further reduces the size considerably. Since the size of these vertex covers is already very small, the $h$-hop vertex cover, for $h > 2$, does not further reduce the size significantly. Thus, we do not present the results for $h > 2$. We also do not report in Table 9 the results of those datasets whose size reduction is less than 20%.

Table 9 also reports the total running time of both $\mu$-reach and $(2, \mu)$-reach, for processing 1 million queries. The result shows that the degradation is acceptable, especially compared with the performance of the other indexes shown in the previous experiments. Thus, for large datasets where the index size is a concern, the $(h, k)$-reach index may offer a possible recourse.

**Table 9: Sizes of vertex cover and 2-hop vertex cover, and total query processing time (elapsed time in milliseconds) of $\mu$-reach and of $(2, \mu)$-reach**

|         | Size of vertex cover | Size of 2-hop vertex cover | Query time of $\mu$-reach | Query time of $(2, \mu)$-reach |
|---------|----------------------|----------------------------|---------------------------|--------------------------------|
| AgroCyc | 389                  | 298                        | 5.56                      | 21.55                          |
| aMaze   | 477                  | 272                        | 14.39                     | 38.70                          |
| Anthra  | 357                  | 278                        | 5.33                      | 21.32                          |
| Ecoo    | 396                  | 302                        | 5.43                      | 21.56                          |
| Kegg    | 618                  | 343                        | 16.12                     | 41.55                          |
| Mtbrv   | 367                  | 287                        | 5.46                      | 21.66                          |
| Nasa    | 1841                 | 1223                       | 18.23                     | 39.48                          |
| Vchocyc | 362                  | 277                        | 5.46                      | 21.71                          |

## 7. RELATED WORK

A large number of indexes have been proposed for processing graph reachability queries [2, 3, 5, 6, 7, 14, 15, 16, 17, 19, 23, 24, 26, 27, 28, 29, 32, 34]. We have analyzed these indexes and discussed why they are not suitable for processing $k$-hop reachability queries in Section 3. We have also discussed why the existing indexes for processing shortest-path queries [1, 13, 17, 30, 31] are not efficient for processing $k$-hop reachability queries in Section 3.5.

Some other variations of graph reachability have also been proposed. For example, Jin et al. [21] studied distance-constraint reachability in uncertain graphs where the existence of an edge is given by a probability, and ask what is the probability that the distance from $s$ to $t$ is less than or equal to a user-defined threshold $d$ in an uncertain graph. Their work focuses on designing probabilistic estimators for estimating the probability of reachability. Jin et al. also proposed constrained graph reachability by requiring edges on the path to have certain labels [20]. There are also many other graph indexes proposed such as for processing subgraph [12] and supergraph [9] queries, which are not applicable for processing reachability queries.

We are also aware of a recent work that applies the concept of vertex cover to construct an index for answering single-source distance queries [8]. They identified the limitation of vertex cover for processing distance queries and proposed a tree-structured index in which every node is a graph that keeps distance information. The $k$-reach graph has a similar limitation, i.e., it is a complete graph, if it is used for processing distance queries. However, the $k$-reach graph for $k$-hop reachability is a significantly smaller sparse graph. Moreover, their index is also too expensive for processing $k$-hop reachability queries.

## 8. CONCLUSIONS

We conclude this paper with a summary of our main contributions as follows.

- To the best of our knowledge, we are the first to study the problem of **$k$-hop reachability**.

- We proposed an efficient index, **$k$-reach**, to process $k$-hop reachability queries. The $k$-reach index is simple in design and easy to implement. In particular, it can effectively handle skewed degree distribution in real-world graphs.

- The $k$-reach index can handle both classic reachability queries (i.e., the case when $k = \infty$) and $k$-hop reachability queries.

- We analyzed the limitations of the existing works in handling $k$-hop reachability (see Section 3). We also showed that even for processing classic reachability queries, $k$-reach is more efficient than the state-of-the-art indexes that are tailored for classic reachability [23, 24, 28, 32].

- Our experimental results also verified the efficiency of $k$-reach in answering $k$-hop reachability queries, for both small and large values of $k$, thus demonstrating its suitability for different real life applications where the value of $k$ may vary. In particular, we showed that $k$-reach is up to orders of magnitude faster than $k$-hop BFS and the shortest-path index [13].

For future work, we plan to study efficient indexing techniques for processing very large graphs, for which the current index may give a large index size and may not scale. I/O-efficient algorithms and parallelization may be necessary to construct indexes for those large graphs, such as the index for shortest paths as studied in [8].

## 9. ACKNOWLEDGMENTS

The authors would like to thank the reviewers for their constructive comments. This research is supported in part by the A*STAR Thematic Strategic Research Programme Grants (102 158 0034 and 112 172 0013), and grants from the Research Grants Council of the Hong Kong SAR (CUHK 411211 and CUHK 419109).